\documentstyle[aps]{revtex}

\begin{document}
\title{Spatial- and spatio-temporal pattern formation in optically driven discrete
systems\bigskip\ }
\author{Victor M.Burlakov\bigskip}
\address{Institute of Spectroscopy Russian Academy of Sciences, 142092 Troitsk,\\
Moscow region, Russia\bigskip\ }
\maketitle

\begin{abstract}
Dynamical coherent structure (pattern) formation in the Klein-Gordon lattice
excited by periodic external field near the optical resonance is studied. It
is shown that besides spatial patterns discovered recently (V.M.Burlakov,
Phys.Rev.Lett. {\bf 80}, 3988 (1998)) the spatio-temporal patterns can be
generated in the lattice in rather broad region of excitation parameters. An
influence of cubic anharmonicity in the lattice potential and that of random
noise on the pattern formation is examined.
\end{abstract}

\section{\bf Introduction}

Dynamical coherent structure (pattern) formation is a new phenomenon in
dynamics of anharmonic lattices. So far patterns have been studied basically
in continuous systems (see Ref.1 and references therein) and recently in
granular materials \cite{patt2,patt3,patt4,patt5}. Some soliton-like
localized structures have been studied in driven anharmonic lattices \cite
{pend_exp,pend_theor1,pend_theor2,rossl1}. Recently it has been found that
optically driven Klein-Gordon (KG) lattice with quartic anharmonicity
possesses pattern formation, i.e. formation of a combination of few standing
waves, or, in the other words, lattice spatial modes (LSMs), vibrating with
one and the same frequency $\omega $ equal to that of the driving field \cite
{BurPrl}. The important factors for the pattern formation are: i) modulation
instability (MI) of the $k=0$ LSM directly excited by external field and
resulting in generation of other LSMs composing the pattern; ii) destructive
interference of modulation instabilities of different LSMs resulting in the
pattern stability. Due to interference of constituent LSMs with each other
the pattern looks like a train of {\it equal} intrinsic localized modes
(ILMs) extensively studied lately \cite
{Dolgov,Sievers,Page,Burl,Page1,Bishop}. Optical creation of a random
ensemble of ILMs have been recently proposed by R\"{o}ssler and Page for
model system \cite{rossl2} and by Lai and Sievers for realistic spin wave
systems \cite{LaiSievers}. Due to equality of the constituent ILMs the
pattern formation opens a new way of experimental identification of ILMs
based on dynamical breaking of the lattice translational symmetry.

To make further step toward experimental identification of patterns (or
ILMs) it seems expedient to study the pattern generation and stability in
optically driven Klein-Gordon (KG) lattice with both cubic and quartic
anharmonicity and in the presence of thermal fluctuations (random noise). In
the present paper I show that the pattern formation conditions in the KG
lattice with cubic and quartic anharmonicity can be analyzed on a bases of
the lattice with pure quartic anharmonicity. The pattern generation and
stability regions in the space of external excitation parameters are
presented. It is shown that at certain conditions random noise results in
frustration of the stationary spatial patterns discovered in \cite{BurPrl}
giving rise to a new type of patterns. The particles vibration amplitude in
the new patterns is modulated in both space and time so one may consider
them as spatio-temporal patterns. The latter can be generated close to the
borders of the spatial patterns stability regions and probably correspond to
a transition regime from spatial patterns to chaos.

\section{MI of optically excited standing wave in the Klein-Gordon lattice}

At the initial stage of the pattern formation the perturbation LSM with $k_p$
is amplified due to MI of the excited $k=0$ nonlinear LSM (referred below as
to {\it excited carrier LSM}, or {\it excited LSM}). Simultaneously the
higher spatial harmonics $2k_p,$ $3k_p,$ ets. will be generated due to
3-wave and/or 4-wave mixing of the excited LSM and the perturbation LSM.
Growing up of the higher harmonics, their mutual interaction and coupling to
the excited LSM finally lead to the pattern formation. Thus, the MI of the
excited LSM works as an ignition for the pattern formation.

Specific feature of the KG lattice is that it possesses a single optical
band of vibrations. The MI of free vibrating waves in various anharmonic
lattices including the KG one was analyzed in \cite
{MI01,MI02,MI03,MI04,MI05,MI06,MI07}. Essential points for analysis of MI in
the driven lattice are those related to i) optical excitation at frequency $%
\omega $ which is different from the eigen frequency of the $k=0$ LSM; ii)
phenomenological damping which mimic all the dissipative interactions
between the phonon band under consideration and any other excitations in
realistic system. Note, that the dissipative interactions which involve only
the vibrations from the optical band of the KG lattice are rigorously
treated in the present analysis.

Motion equation for $n-th$ particle of unity mass and charge in the KG
lattice is

\begin{equation}
d^2U_n/dt^2+\gamma \cdot dU_n/dt+\omega _0^2U_n+K_2\cdot \left(
2U_n-U_{n-1}-U_{n+1}\right) +K_3U_n^2+K_4U_n^3=E_0e^{i\omega t}+c.c.,
\label{eq1}
\end{equation}
where $\gamma $ is phenomenological damping constant, $\omega _0^2$, $K_4$
are incite and $K_2$ is intercite force constants, $E_0$ is the external
field amplitude. The trial solution of Eq.(\ref{eq1}) for small
anharmonicity can be chosen in the form

\begin{equation}
U_n(t)=\frac 12\left[ V_{C0}+V_{C1}\exp \left( i\omega t+i\varphi _1\right)
+V_{C2}\exp \left( i2\omega t+i\varphi _2\right) +c.c.\right] ,  \label{eq2}
\end{equation}
where $V_{Cj}$ and $\varphi $ are real amplitude and phase angle
respectively. Two types of the KG lattice are considered. The lattice with
pure quartic anharmonicity (L1): $\omega _0=354,$ $K_2=6\cdot 10^4,$ $K_3=0,$
$K_4=-3.3\cdot 10^6;$ the lattice with both cubic and quartic anharmonicity
(L2): $\omega _0=354,$ $K_2=6\cdot 10^4,$ $K_3=6\cdot 10^5,$ $K_4=5\cdot
10^5 $. The incite potential curves for a particle in the L1 and L2 lattices
are shown in Fig.1a. A particle vibration in the right potential well is to
be considered. One can see from Fig.1a that the particle remains in the
minimum until its vibration amplitude do not exceed the value $%
U_{th}^0\simeq 0.2$. The latter value can be considered as a natural
threshold for the point defect formation related to the out-off minimum jump
of the particle. The parameters of the L1 and L2 lattices were chosen in a
way for the $V_{C1}\propto E_0$ dependences to be nearly the same (see
Fig.1b). This choice allows to understand a pure influence of cubic
anharmonicity not related to anharmonic shift of the optical mode frequency.
For $\omega <\omega _0$ the $V_{Cj}\propto E_0$ dependences show bistability
and for $\omega >\omega _0$ these dependences are monotonous. Below only the
latter case is considered in detail.

Obviously $V_{C0}=V_{C2}=0$ in the L1 lattice and one may use the rotating
wave approximation (RWA) for stability analysis of the excited LSM.
According to Ref. \cite{BurPrl} the perturbation growing rate (increment) $%
\mathop{\rm Im}
(\Omega (q))$ can be determined from the equation 
\begin{equation}
\left[ \omega _q^2-\left( \omega +\Omega \right) ^2-i\gamma \cdot \left(
\omega +\Omega \right) \right] \cdot \left[ \omega _q^2-\left( \omega
-\Omega \right) ^2+i\gamma \cdot \left( \omega -\Omega \right) \right]
=\beta ^2,  \label{eq4}
\end{equation}
obtained after substitution of (\ref{eq2}) with perturbation 
\begin{equation}
\delta U_n=\frac 12\cos (qn)\cdot \left[ V_{P1}\exp (i\cdot (\omega -\Omega
)t)+V_{P2}\exp (i\cdot (-\omega -\Omega )t)+c.c.\right] ,  \label{eq4a}
\end{equation}
into Eq(\ref{eq1}) (here and below the wavevector of perturbation LSM is
denoted by $q$ while that of carrier LSM is denoted by $k$). Here $\omega
_q^2=\omega _0^2+4K_2\sin (q/2)^2+\frac 32K_4V_{C1}^2,$ $\beta =\frac 34%
K_4V_{C1}^2$, $V_{Pj}$ is the complex amplitude, and $q$ and $\Omega $ are
the wave vector and the complex frequency shift of the perturbation wave
respectively.

To study MI of the solution (\ref{eq2}) in general case the perturbation has
to be chosen in the form 
\begin{equation}
\begin{array}{c}
\delta U_n=\frac 12\cos (qn)\cdot \left[ V_{P0}\exp (-i\Omega t)+V_{P1}\exp
(i\cdot (\omega -\Omega )t)+V_{P2}\exp (i\cdot (-\omega -\Omega )t)+\right.
\\ 
\left. V_{P3}\exp (i\cdot (2\omega -\Omega )t)+V_{P4}\exp (i\cdot (-2\omega
-\Omega )t)+c.c.\right] .
\end{array}
\label{eq3}
\end{equation}
The $%
\mathop{\rm Im}
(\Omega (q))$ function then was calculated numerically equating to zero the
determinant of the corresponding system of five liner equations. Relative
increment $%
\mathop{\rm Im}
(\Omega (q)/\omega )$ for the L1 and L2 lattices and that for free carrier
LSM with $k=0$ in the L2 lattice with $\gamma =0$ are plotted in Figs.2a-c
respectively for $\omega =1.05\omega _0$ and three values of $V_{C1}$. One
can see that the instability regions ($%
\mathop{\rm Im}
(\Omega (q))>0$) in the $q$ space for the excited carrier LSM in the L1
lattice are much broader and the instability is stronger than those in the
L2 lattice. Note that the instability of free carrier LSM is much stronger
than that of excited one (compare Fig.2b and Fig.2c). By symbols in Fig.2b
are shown the $%
\mathop{\rm Im}
(\Omega (q)/\omega )$ curves for the system with pure quartic anharmonicity
which will be used for approximation of the L2 lattice. The effective
quartic anharmonicity constant was chosen to be $K_4^{eff}=K_4\cdot \left(
1+\left( \frac{K_3}{\omega _0^2}\right) ^2\cdot V_{C1}^2\right) -\frac{11}{12%
}\left( \frac{K_3}{\omega _0}\right) ^2$ and also the factor $\beta $ in (%
\ref{eq4}) was changed in this case to $\beta ^{eff}=\frac 7{12}K_4V_{C1}^2$%
. The numerical factors $\frac{11}{12}$ and $\frac 7{12}$ were chosen for
the best fit of the increment curves shown by lines in Fig.2b. One can see
that the system with the effective quartic anharmonicity approximates rather
well the original L2 lattice if the excitation field $E_0$ is not very high.

This approximation allows to suitably represent the excited LSM instability
regions in the $\left( E,\gamma \right) $ space. The MI regions of the
excited LSM for the perturbation wave vectors $q=\pi /3$ and $q=\pi /2$ are
shown in Fig.3 by solid lines. The curves were obtained from Eq.(\ref{eq4})
under conditions $\Omega (q=\pi /3)=0$ and $\Omega (q=\pi /2)=0$ for the
left and right closed regions respectively. Additionally, the excited LSM
amplitude must be restricted by a value much lower than the $U_{th}^0$. This
is because the vibration amplitude of some particles in the pattern
increases compared to its value in the excited LSM and may result in the
out-of minimum jump of these particles. In the case shown in Fig.3 the
excited LSM amplitude was restricted by the value $0.13\cdot a$, i.e. about $%
0.7\cdot U_{th}^0$, what cuts the MI regions for high $E$ values. In fact,
the difference between particles vibration amplitudes in the excited LSM on
one hand and in the pattern on the other hand depends on damping so the MI
regions can be more extended to the higher $E$ values at high $\gamma $
values.

\section{Pattern solutions for $K_3=0$ and their stability}

The perfect pattern can be generated starting from the seeding LSM $\sim
\cos (q\cdot n)$ at $t=0$. Then under action of external field $%
E=E_0e^{i\omega t}$ the system will pass through two stages: a) excitation
of carrier LSM; b) growing up of the seeding LSM due to MI of the excited
carrier LSM and simultaneous generation of other LSMs due to four-wave
mixing. Obviously, if the number of generated LSMs at the initial stage of
the pattern formation is fairly restricted ($3\div 5)$ the LSMs may grow up
to large enough amplitude to be as additional nonlinear waves (carrier LSMs)
in the lattice and to stabilize the pattern. We restrict our consideration
by the total number of carrier LSMs $N_{LSM}=3$ and $N_{LSM}=4$ (3-LSM- and
4-LSM patterns, respectively). The former is build up of the carrier LSMs
with wave vectors $k_1=0,$ $k_2=\pi /2$ and $k_3=\pi $ (lattice constant $%
a=1 $) while the latter with those $k_1=0,$ $k_2=\pi /3,$ $k_3=2\pi /3$ and $%
k_4=\pi $. The MI of the excited carrier LSM must have a maximum around $%
q_{\max }=\pi /2$ in case of $N_{LSM}=3$ and around $q_{\max }=\pi /3$ for $%
N_{LSM}=4$. This is the condition for $E_0$ or, in the other words, for the
excited carrier LSM amplitude $V_C$. For $N_{LSM}=2$ ($q_{\max }=\pi $) no
solutions have been found.

Because of the symmetry arguments the pattern may consist of standing LSMs
only. The trial pattern solutions are 
\begin{equation}
\begin{array}{c}
U_n^{3-LSM}(t)=\frac 12\left[ V_{C1}\exp \left( i\omega t+i\varphi _1\right)
+V_{C2}\cos (\frac \pi 2n)\exp (i\omega t+i\varphi _2)+\right. \\ 
\left. V_{C3}\cos (\pi n)\exp (i\omega t+i\varphi _3)+c.c.\right] ,
\end{array}
\label{eq6a}
\end{equation}
\begin{equation}
\begin{array}{c}
U_n^{4-LSM}(t)=\frac 12\left[ V_{C1}\exp \left( i\omega t+i\varphi _1\right)
+V_{C2}\cos (\frac \pi 3n)\exp (i\omega t+i\varphi _2)+\right. \\ 
\left. V_{C3}\cos (\frac 23\pi n)\exp (i\omega t+i\varphi _3)+V_{C4}\cos
(\pi n)\exp (i\omega t+i\varphi _4)+c.c.\right] ,
\end{array}
\label{eq6b}
\end{equation}
for 3-LSM and 4-LSM patterns respectively. No new LSMs important within RWA
will appear due to four-wave mixing of those chosen. Here again $V_{Cj}$ are
real amplitudes and $\varphi _j$ are phase angles. According to $\omega $
and $\gamma $ values a single stable nontrivial (all $V_{Cj}\neq 0$)
solution $\Phi _1$ of the form (\ref{eq6a})-(\ref{eq6b}) was found in the
region between the dotted curves in Fig.3. Outside this region the solution $%
\Phi _1$ is unstable in the sense discussed below. An additional and
strongly unstable solution $\Phi _2$ \cite{BurPrl} exists for the 3-LSM
pattern at $\gamma <0.1\cdot \omega _0$ while for 4-LSM patterns no
additional solutions have been found. The examples of the 3- and 4-LSM
patterns of the $\Phi _1$-type calculated after substitution of (\ref{eq6a})
and (\ref{eq6b}) respectively into Eq.(\ref{eq1}) with the parameters
corresponding to the points {\bf a} and {\bf c} in Fig.3 are presented in
Fig.4. Note, that indeed the patterns can be regarded as the lattices of
intrinsic localized vibrations of the odd parity \cite{Dolgov,Sievers}.

Linear stability analysis of the solution (\ref{eq6a}) within RWA was given
in \cite{BurPrl}. Similar approach to the 4-LSM pattern stability requires
the total perturbation to contain all perturbation waves coupled to each
other via four-wave mixing, i.e. all spatial harmonics resulting from a
product of any three carrier LSMs from Eq.(\ref{eq6b}) on a perturbation
wave. One can see that a set of waves

\begin{equation}
\begin{array}{c}
\delta U_n=\frac 12\left\{ \exp (i\cdot (\omega -\Omega )t)\cdot \left[
V_{P1}\cos (qn)+V_{P3}\cos ((\frac \pi 3-q)n)+V_{P5}\cos ((\frac \pi 3%
+q)n)+\right. \right. \\ 
\left. V_{P7}\cos ((\frac{2\pi }3-q)n)+V_{P9}\cos ((\frac{2\pi }3%
+q)n)+V_{P11}\cos ((\pi -q)n)\right] + \\ 
\exp (i\cdot (-\omega -\Omega )t)\cdot \left[ V_{P2}\cos (qn)+V_{P4}\cos ((%
\frac \pi 3-q)n)+V_{P6}\cos ((\frac \pi 3+q)n)\right. \\ 
\left. \left. +V_{P8}\cos ((\frac{2\pi }3-q)n)+V_{P10}\cos ((\frac{2\pi }3%
+q)n)+V_{P12}\cos ((\pi -q)n)+c.c.\right] \right\}
\end{array}
\label{eq7}
\end{equation}
fulfills this condition. Indeed the set of waves (\ref{eq7}) contains
spatial harmonics $k_p=\pm q+\frac \pi 3m$ ($m=0,$ $1,$ $2,3$). After
coupling to any three carrier LSMs from (\ref{eq6b}) it results in a set of
spatial harmonics with wave vectors $k_p=\pm q+\frac \pi 3m\pm \frac \pi 3l$
($l=0,1,2,3$) which obviously can be reduced to (\ref{eq7}). System of
twelve linear equations derived after substitution of (\ref{eq6b}) and (\ref
{eq7}) into Eq.(\ref{eq1}) was solved numerically to determine $\Omega (q).$

The perturbation growing rate $%
\mathop{\rm Im}
\left( \Omega \right) $ versus wavevector is shown in Fig.5 for 4-LSM
patterns generated at the points {\bf a} and {\bf b} in Fig.3. The pattern
(a) in Fig.5 is obviously stable, i.e. $%
\mathop{\rm Im}
\left( \Omega (q)\right) <0$ for $0\leq q\leq \pi $, while the pattern (b)
is unstable ($%
\mathop{\rm Im}
\left( \Omega (q)\right) >0$ for some $q$ values). As some individual
carrier LSMs composing the pattern are unstable the pattern stability can
come from destructive interference between MIs of the carrier LSMs. To
qualitatively understand this phenomenon one may imagine the simplified
picture: i) perturbation LSMs of different symmetry compose the perturbation
state Eq. (\ref{eq7}) via four-wave mixing; ii) this state is excited by
different (by symmetry) coherent excitations (different carrier LSMs).
Hence, whether the perturbation state will be continuously populated depends
on the amplitudes of the coherent excitations and on the phase angle(s)
between them (the influence of phase angle between the carrier LSMs on the
pattern stability is demonstrated in Fig.5). It may happen that at certain
combination of the excitation parameters the perturbation state won't be
populaed at all. That means the MI-mediated influence of a given carrier LSM
on the perturbation (\ref{eq7}) can be cancelled by that of the other(s)
resulting in the stability of all the carrier LSMs, i.e. the pattern.

\section{Pattern formation in the presence of noise}

The stability regions for both 3-LSM and 4-LSM patterns corresponding to the
condition $%
\mathop{\rm Im}
\left( \Omega \right) <0$ (damping of the perturbation) are shown by dotted
lines in Fig.3. These stability regions were determined numerically via
excitation of the pattern starting from a small amplitude seeding wave
(S-patterns). Since the stability regions are inside the MI regions of the
corresponding excited carrier LSM one might expect that the stable patterns
can be spontaneously generated there. This is not always true, however, if
the pattern is generated without any seed but in the presence of random
noise (N-pattern). When generated from a small amplitude noise the pattern
can be unstable because of simultaneous generation of waves with wrong ($\pi
/q$ is not integer) wavevectors. The examples of the N-pattern formation are
given in Fig.6. Numerical study was performed for 60 particle chain using
the standard conservative scheme of numerical integration of the motion
equations (\ref{eq1}) with cyclic boundary conditions. Random noise of the
value between $\pm 0.00005a$ was generated at each time step and added to
the current particles positions. The panels notation in Fig.6 correspond to
the points notation in Fig.3. The 4-and 3-LSM patterns generated
respectively at the points {\bf a} and {\bf c} in Fig.3 are really stable
and can be attributed to the true spatial patterns since they are similar to
those calculated theoretically (compare dotted and solid lines in Fig.4).
The pattern at the point {\bf d} is inside the 3-LSM stability region and
shows periodic modulation in time similar to that of the pattern at the
point {\bf b }outside the corresponding stability region. Therefore the
N-patterns at the points {\bf b} and {\bf d} can be regarded as
spatio-temporal patterns rather than the true spatial patterns.

Spatio-temporal patterns probably also possess certain stability properties
though the latter are not so definite as for the spatial patterns. Typical
temporal evolution of the spatio-temporal pattern is shown in Fig.7.
According to Fig.7 the pattern corresponding to the point {\bf b} in Fig.3
is formed over $\sim 30$ periods of vibration and remains rather stable
though its contrast slightly decreases in time. Relatively long-time
stability of the pattern outside the stability region of the true spatial
pattern can be qualitatively understood. Growing up of a perturbation $%
\delta U_n(t)$ in the presence of the unstable pattern at the point {\bf b}
results in the shift of the quasiharmonic resonance frequency of the optic
LSM 
\[
\Delta \omega (k=0)^2=3K_4\cdot \left\langle \delta U_n^2(t)\right\rangle
_{n,t} 
\]
and consequently in the increase of the frequency mismatch $\omega -\omega
(k=0)$. Due to this increase the pattern approaches stability region or even
enters it and the growing conditions for the perturbation are violated.
After damping of the perturbation the pattern returns to the initial point
and the process starts again thus giving rise to slow modulation of the
pattern amplitude, i.e. results in formation of spatio-temporal pattern. The
latter can be stable if the pattern modulation is perfectly periodic, i.e.
the modulation amplitude doesn't grow up, otherwise it is obviously
unstable. In the long-time scale the evolution of the unstable
spatio-temporal pattern due to continues energy collapse can lead to a
situation when the vibration amplitude of a particle exceeds the threshold
value $U_{th}^0$ what means the point defect formation.

\section{Optical intensity required for the pattern generation}

Our numerical experiments show that to reach the threshold for the pattern
formation with $N_{LSM}=4$ in the L1 lattice the particles in the excited
LSM must vibrate with the amplitude $A_p\simeq 0.05a$. The electric field
strength $E_0$ of the optical excitation at $\omega =1.05\omega _0\simeq
10^2 $ $cm^{-1}$ can be then estimated using motion equation in the harmonic
approximation. Suggesting $a=5\AA $ and $\gamma =0.05\omega _0$ one obtains $%
E_0=A_pm_pm_e[(\omega _0^2-\omega ^2)^2+(\gamma \omega
)^2]^{1/2}/(e_pe_e)\simeq m_p/e_p$ $[{\rm {V/cm}],}$ where $m_p$ and $e_p$
are the particle mass and charge measured in the free electron units $m_e$
and $e_e$ respectively. Accordingly, the field strength is of the order of $%
1\,{\rm {V/cm}}$ for light particles like electrons and of the order of $%
10^5\,{\rm {V/cm}}$ for ionic solids. The latter value of $E_0$ can be
reached in the laser pulses. Obviously the pulse duration $T$ must be enough
for the pattern to be generated and detected. The generation stage lasts
over $30-40$ periods of vibration and $10-20$ periods are probably needed
for the pattern detection what in total means $T\simeq 20$ $ps$. Hence, for
the pattern formation in an electronic system (e.g. charge-density wave
conductor) the IR laser pulses of energy $W\simeq 10^{-11}$ $mJ$ focused
into $\sim 0.01$ $cm^2$ are required. The pattern formation in an ionic
system needs much higher pulse energy $W\simeq 0.1$ $mJ$ unless the smaller
focusing area is used.

Thus, in real physical experiment a system will be subjected to external
field and examined during a finite time. Therefore it seems reasonable to
use some criterion of the pattern {\it relative} stability (during finite
time period) rather than that given by linear stability analysis. The
latter, moreover, seems to be too complicated for spatio-temporal patterns
especially for a system with realistic potential. According to the above
said the pattern stability regions in Fig.3 must be transformed taking into
account the pattern generation and probably detection conditions (see next
Section).

\section{Pattern generation and stability for $K_3\neq 0$}

The 3-LSM pattern solution in this case have the form 
\begin{equation}
\begin{array}{c}
U_n^{3-LSM}(t)=\frac 12\left[ V_{C10}+V_{C11}\exp \left( i\omega t+i\varphi
_{11}\right) +V_{C12}\exp \left( i2\omega t+i\varphi _{12}\right) +\right.
\\ 
\cos (\frac \pi 2n)\cdot \left( V_{C20}+V_{C21}\exp \left( i\omega
t+i\varphi _{21}\right) +V_{C22}\exp \left( i2\omega t+i\varphi _{22}\right)
\right) + \\ 
\left. \cos (\pi n)\cdot \left( V_{C30}+V_{C31}\exp \left( i\omega
t+i\varphi _{31}\right) +V_{C32}\exp \left( i2\omega t+i\varphi _{32}\right)
\right) +c.c.\right] .
\end{array}
\label{eq9}
\end{equation}
In contrast to Eqs (\ref{eq6a}) and (\ref{eq6b}) the pattern (\ref{eq9})
contains static distortion and also the higher (second) temporal harmonic
can not be neglected. Hence the linear stability analysis of the pattern in
case $K_3\neq 0$ is too complicated. Note, that the approximation of the
system with $K_3\neq 0$ by a system with effective quartic anharmonicity
gives reasonably good results when applied to the carrier LSM stability but
not to the pattern stability. According to the estimation given in the
previous Section, one can revise the pattern stability conditions and
consider the pattern as being relatively stable if it is stable during few
tens of periods $T=2\pi /\omega $. As a new stability criterion one may
consider a requirement for a pattern to be stable if it shows well
pronounced spatial modulation (%
\mbox{$>$}
10 per cent) of particle vibration amplitude for a time interval of about
30-50 periods of vibration. The latter time for real system is determined by
the pattern detection time and can be even shorter.

The new regions of the pattern relative stability were determined from
numerical experiment on the bases of the aforementioned criterion (see
shaded regions in Fig.8). The bubbles in Fig.8 show the pattern generation
(excited LSM instability) regions for the L1\ and L2 lattices (solid and
dotted lines, respectively). Dashed-dotted thin lines denote the envelope
for the corresponding bubbles and determine the MI cut-off for the excited
carrier LSMs. These cut-off lines are determined from the system of
equations 
\begin{equation}
\left\{ 
\begin{array}{c}
\Omega (q)=0, \\ 
\frac{\partial \Omega (q)}{\partial q}=0, \\ 
E=\sqrt{\left( \omega _0^2+\frac 34K_4\cdot V_{c1}^2-\omega ^2\right)
^2+\left( \gamma \omega \right) ^2},
\end{array}
\right.  \label{eq10}
\end{equation}
where $\Omega (q)$ is determined by Eq. (\ref{eq4}) and again $K_4$ and $%
\beta $ must be substituted with $K_4^{eff}$ and $\beta ^{eff}$ if $K_3\neq
0 $. Note, that the stability regions are separated from the envelope lines
because in the close vicinity to the latter the perturbation growing rate is
too low to form the pattern with well pronounced modulation (at least 10 per
cent) during the time of numerical experiment.

It is important to determine the pattern generation and relative stability
regions in the $\left( E,\omega \right) $ space, i.e. the space of external
excitation parameters. They are shown by shaded regions in Fig.9 for 3- and
4-LSM patterns in the L2 lattice with $\gamma =0.025\cdot \omega _0$. The
latter value seems quite reasonable for realistic systems. The stability
regions were obtained numerically under condition that the pattern shows
more than 10 percent modulation in $\left\langle U_n^2(t)\right\rangle _t$
function but doesn't result in the point defect formation (particle jump
from its initial minimum in Fig.1a) during 120 periods of vibration. Even in
the case of nearly fastest evolution of the 3-LSM pattern it can be
considered as stable for $\sim 10$ periods of vibration (see Fig.10) though
in the long time scale the system eventually arrives to chaotic state.
Taking into account these short-living patterns one may expect the pattern
formation phenomenon in rather broad region of the external excitation and
system parameters. Note, that the particles vibration in the time interval $%
3-4$ (90-120 periods) in Fig.10 is similar to that in the spatio-temporal
pattern in Fig.6c suggesting that the spatio-temporal patterns represent the
intermediate state in transition from dynamical order to chaos.

\section{Conclusions}

The dynamical coherent structure (pattern) formation was studied in the
optically driven Klein-Gordon lattice with cubic and quartic anharmonicity
in the presence of random noise. It is shown that in general the patterns
are characterized by both spatial and temporal modulation of the particles
vibration amplitude (so-called spatio-temporal patterns) and can be
relatively stable in rather broad region of external excitation and the
system parameters.

\section{Acknowledgments}

This work was supported by Russian Ministry of Science within the program
''Fundamental Spectroscopy''.

\newpage\ 

\begin{center}
{\bf Figure captures}
\end{center}

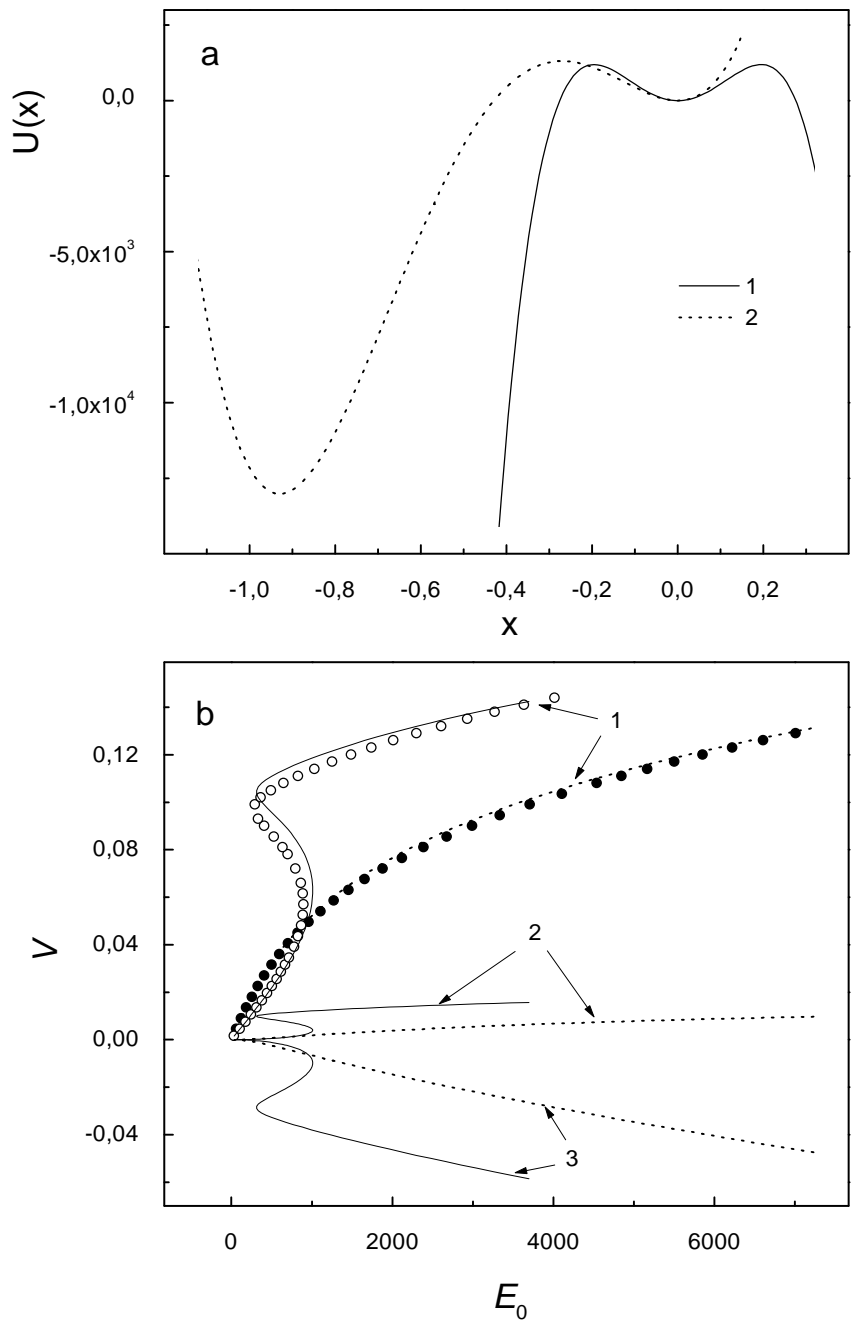
Fig.1. a) Incite potential function for a particle in the Klein-Gordon L1-
(1) and L2 (2) lattices (see the text).

b) $V_{Cj}\propto E_0$ plot (see Eq. 2 in the text). Curves 1, 2 and 3
correspond to $j=1$ (first harmonic)$,2$ (second harmonic) and $0$ (static
displacement or, zero harmonic) respectively. Lines correspond to potential
of the L1 lattice and symbols to that of the L2 lattice. The external field
frequency is $\omega =0.9\omega _0$ (solid lines) and $\omega =1.05\omega _0$
(dotted lines), $\gamma =0.025\omega _0$.

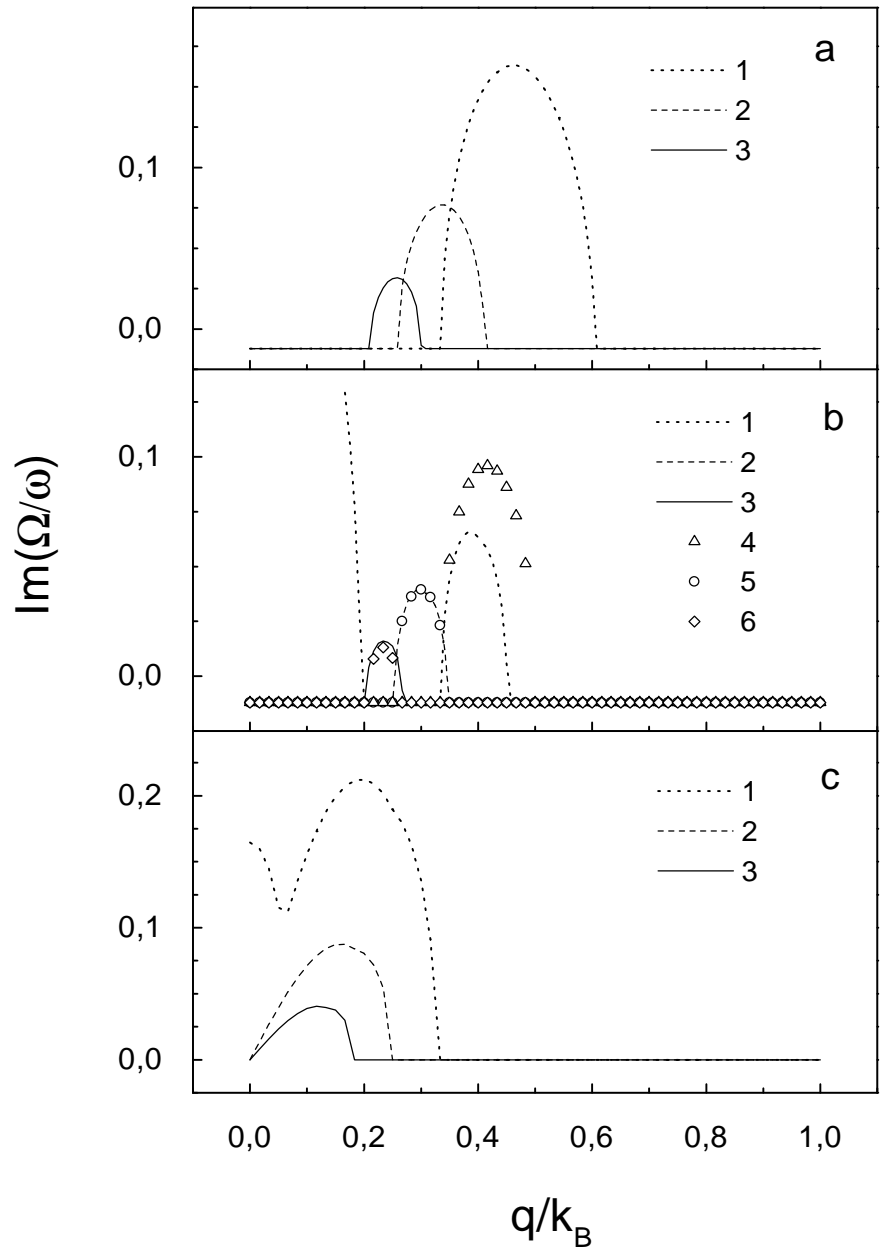
Fig.2. Perturbation relative growing rate (relative increment) $%
\mathop{\rm Im}
(\Omega (q)/\omega )$ for $\omega =1.05\omega _0$ and $\gamma =0.025\omega
_0 $ in the L1- (a) and L2 (b) lattices; (c) MI of the free ($\gamma =0,$ $%
E_0=0 $) nonlinear LSM with $k=0$ in the lattice L2. External field
amplitude $E_0$ was chosen for panels a) and b) in a way when the $V_{C1}$
value is equal to $0.14$ (1), $0.1$ (2) and $0.07$ (3). Curves (4)-(6) in
the panel (b) were calculated from Eq. (\ref{eq4}) for the lattice of
L1-type with effective quartic anharmonicity $K_4^{eff}=K_4\cdot \left(
1+\left( \frac{K_3}{\omega _0^2}\right) ^2\cdot V_{C1}^2\right) -\frac{11}{12%
}\left( \frac{K_3}{\omega _0}\right) ^2$ and $\beta ^{eff}=\frac 7{12}%
K_4V_{C1}^2$ (see text). The $E_0 $ values for curves (4), (5) and (6) in
the panel b) are equal to those corresponding to $V_{C1}$ values for curves
(1), (2) and (3) respectively.

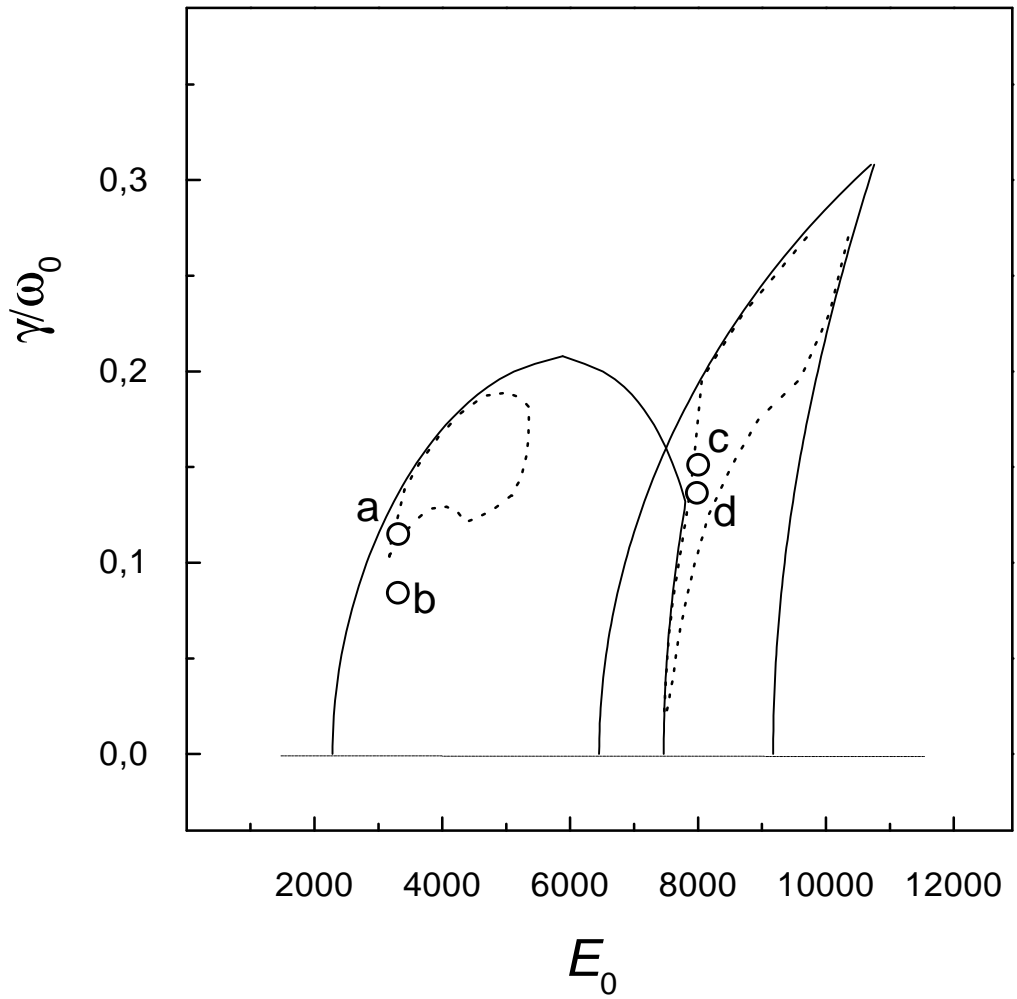
Fig.3. Carrier LSM modulation instability regions in the $(E_0,\gamma )$
space enclosed by solid curves and corresponding to the pattern generation
by external field with $\omega =1.1\omega _0$ (3-LSM patterns) and with $%
\omega =1.05\omega _0$ (4-LSM patterns) in the L1 lattice. The calculated
pattern stability regions are enclosed by dotted curves.

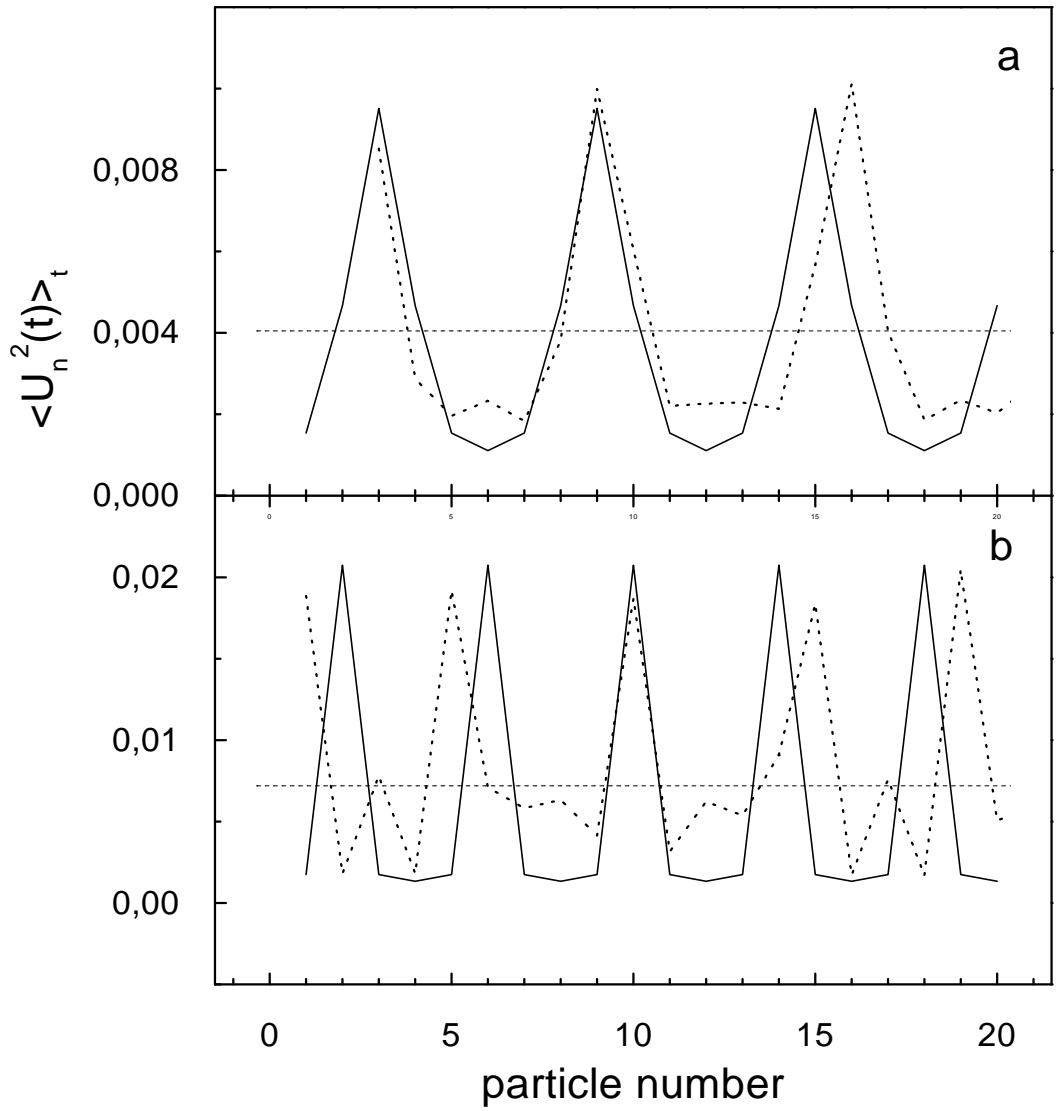
Fig.4. The stable 3-LSM and 4-LSM patterns (the $<U_n^2(t)>_t$function)
calculated (solid lines) and obtained by numerical simulation in the
presence of noise (dotted lines) at the points {\bf a} (a) and {\bf b} (b)
in Fig.3. Thin dashed lines show the amplitude square of the carrier LSM
prior the pattern formation has began.

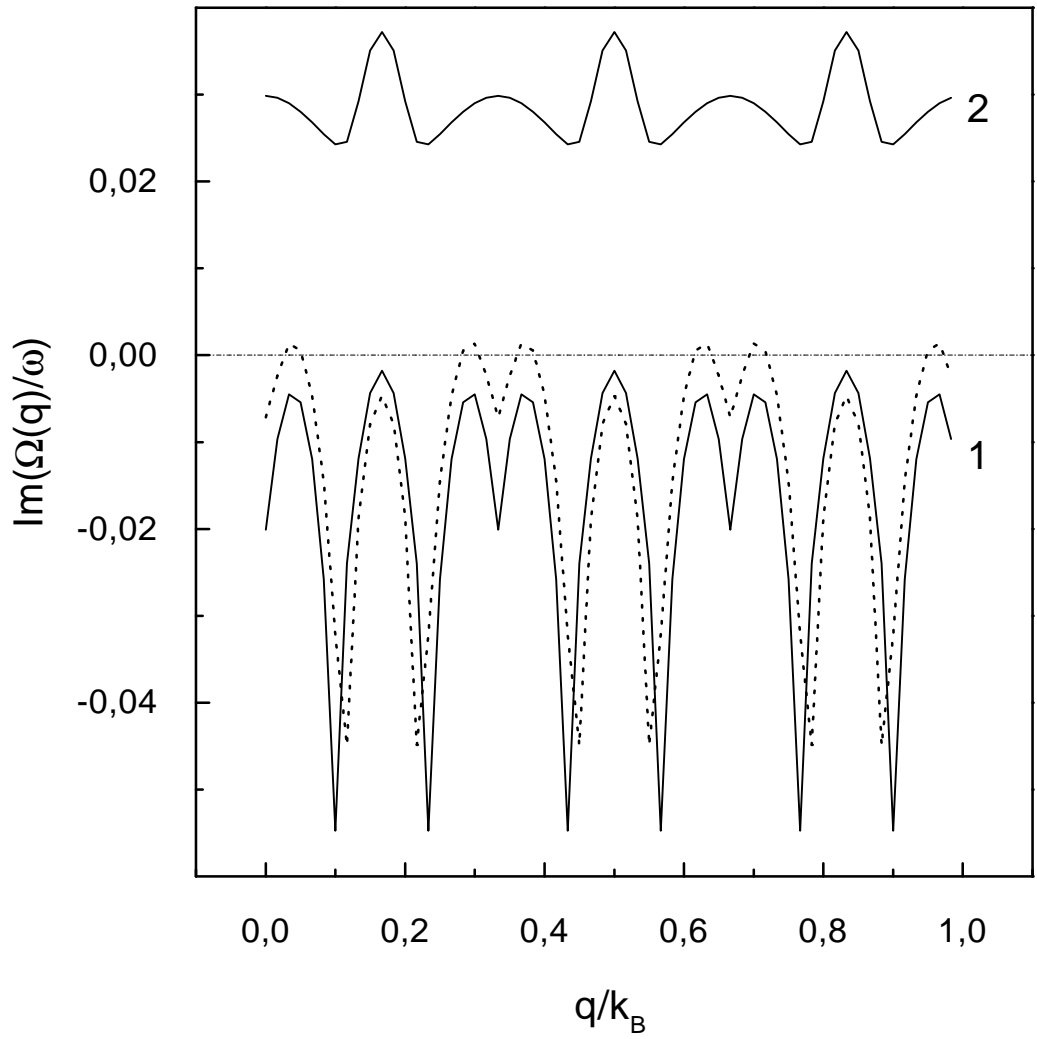
Fig.5. Relative increment $%
\mathop{\rm Im}
(\Omega (q)/\omega )$ calculated for stable (1) and unstable (2) 4-LSM
patterns (\ref{eq6b}) corresponding to the points {\bf a} and {\bf b} in
Fig.3. Dotted line shows the $%
\mathop{\rm Im}
(\Omega (q)/\omega )$ function calculated with the same parameters as (1)
but with $\phi _1$ substituted with $\phi _1+\pi /4$.

Fig.6a-d. Molecular dynamic simulation of the N-pattern (see text) formation
in 60-particle L1 lattice with cyclic boundary conditions for $\gamma $ and $%
E_0$ corresponding to the points {\bf a}-{\bf d} respectively in Fig.3.
Particle displacements from corresponding equilibrium positions are
multiplied by factor 3 for (a) and (b) and by factor 2 for (c) and (d).

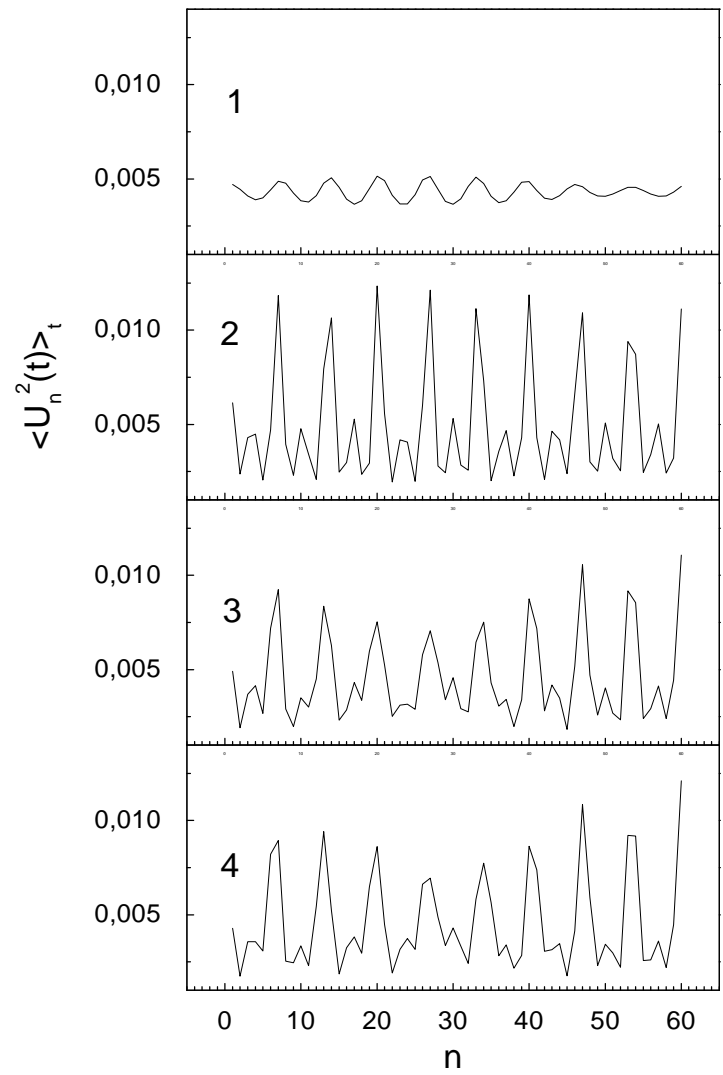
Fig.7. The $<U_n^2(t)>_t$ pattern obtained from numerical experiment with L1
lattice at the point {\bf b} in Fig.3 and taken after 25 (1), 50 (2), 75
(3), and 100 (4) periods of external field vibration. The time average was
taken over 10 periods.

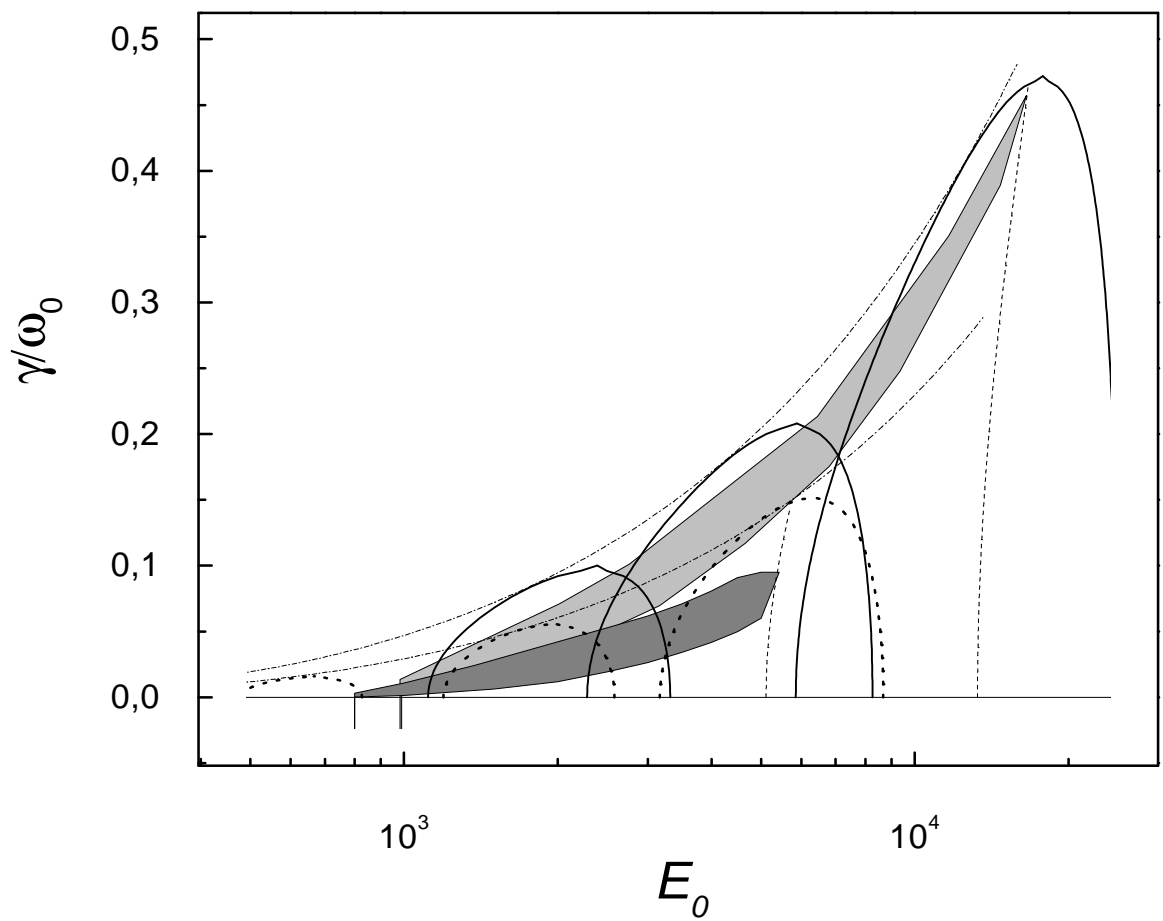
Fig.8. The pattern generation and relative stability regions (shaded) for
the L1 lattice (light grey) and the L2 lattice (dark grey) for $\omega
=1.05\omega _0$. The bubbles enclose the 5-, 4- and 3-LSM pattern generation
regions in the L1- (light grey) and L2 (dark grey) lattices. Dashed lines
denote the envelope for the corresponding bubbles.

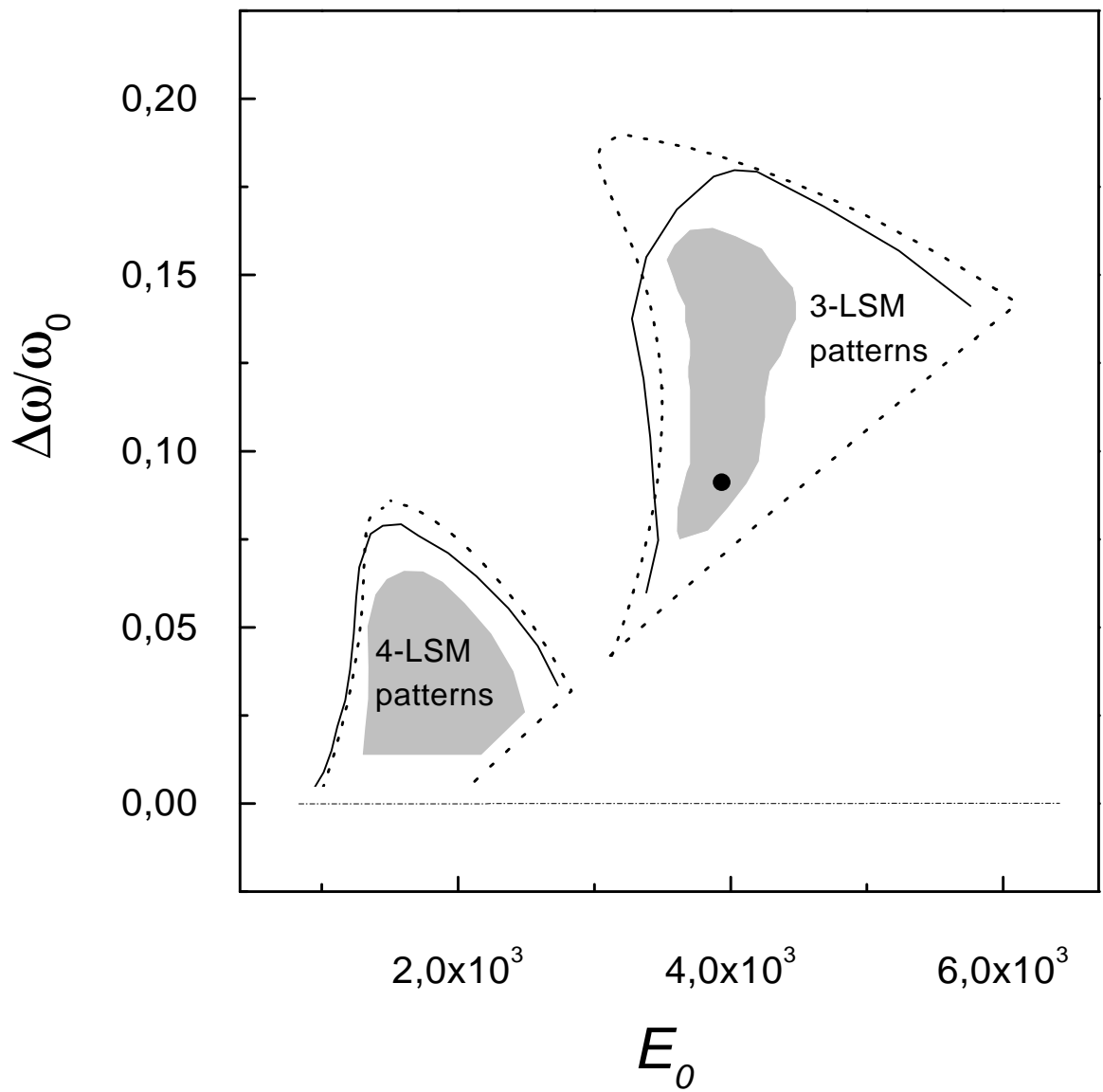
Fig.9. The pattern generation and relative stability regions (shaded) for
the L2 lattice with $K_2=30000,$ $\gamma =0.025\omega _0$. Solid lines
enclose the pattern generation regions determined via numerical simulation
for the L2 lattice while dotted lines enclose those obtained from equations $%
\Omega (q=\pi /3)=0$ and $\Omega (q=\pi /2)=0$ for the L1 lattice with
effective $K_4$ and $\beta $ (see capture to Fig.2). In the calculations the
threshold value for the carrier LSM amplitude was taken equal to $\simeq
0.1\cdot a$. Solid symbol denotes the point at which the pattern formation
process is presented in Fig.10.

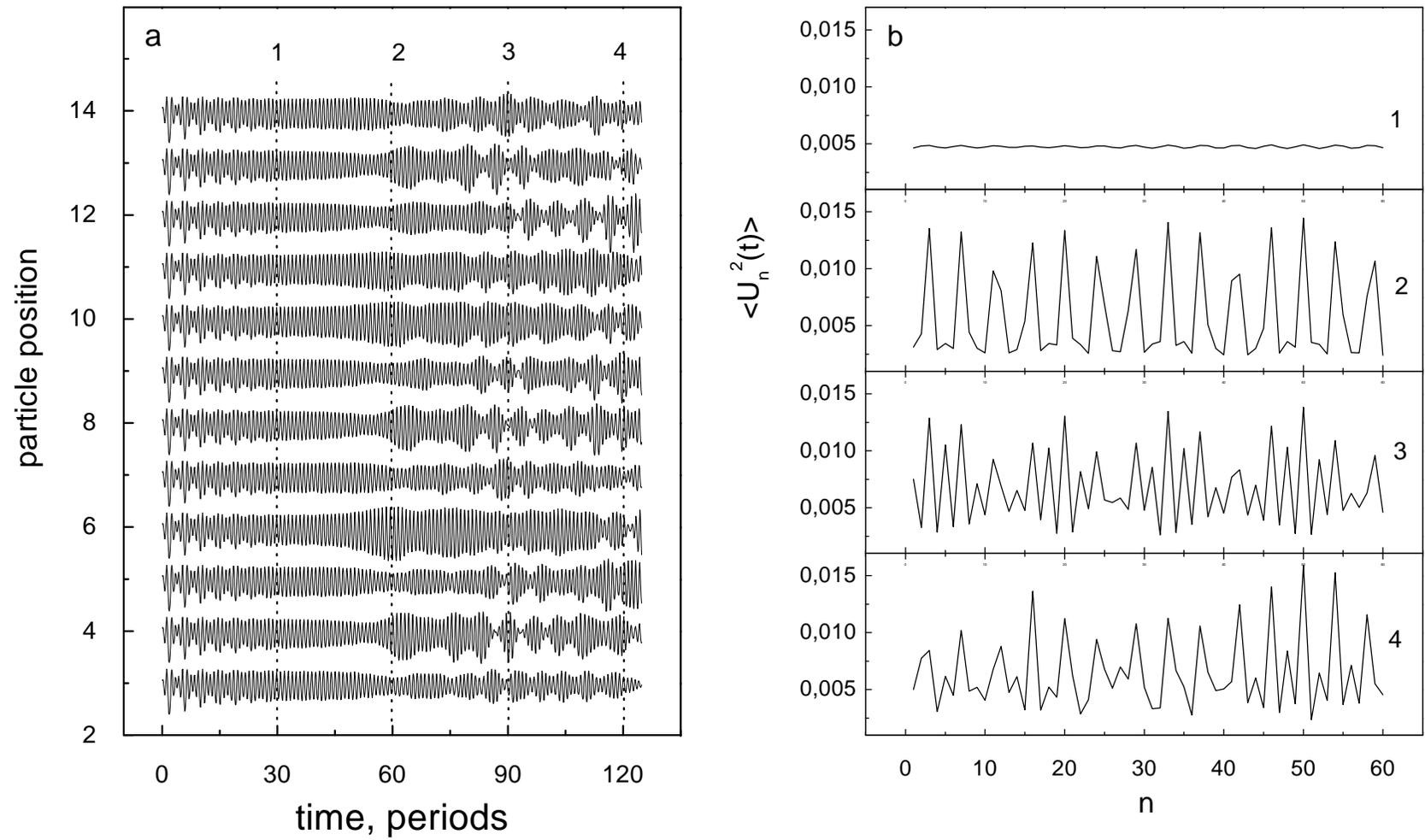
Fig.10. a) Molecular dynamic simulation of the N-pattern formation in the
60-particle L2 lattice for external field parameters corresponding to the
point shown in Fig.9. Particle displacements from corresponding equilibrium
positions are multiplied by factor 3. Vertical dotted lines denote the
starting time moments of the $<U_n^2(t)>_t$ calculation shown in b). The
averaging was taken over 10 periods.

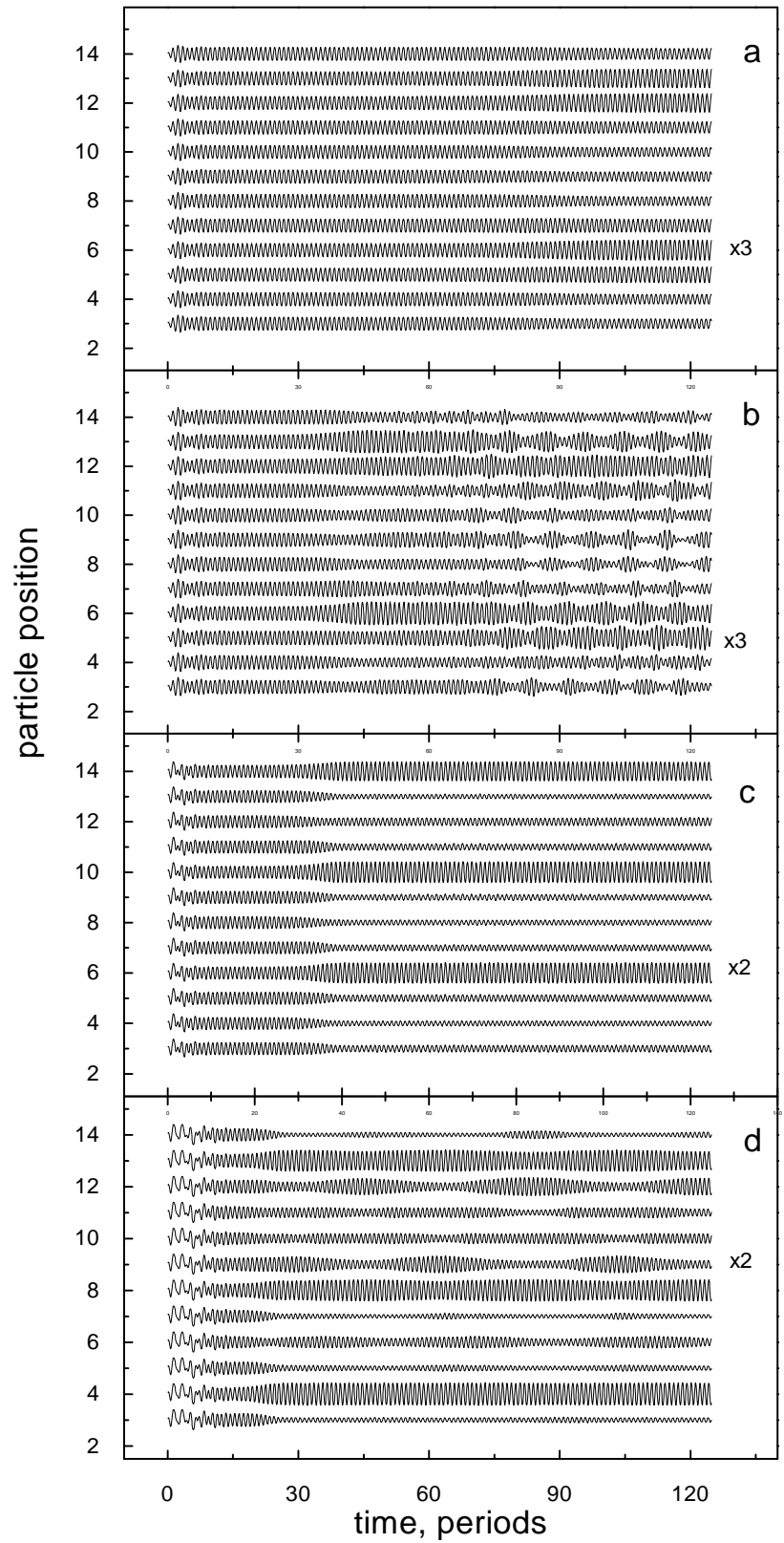
\end{document}